\documentclass[twocolumn]{aastex63}

\newcommand{\kms}{km s$^{-1}$}
\newcommand{\rbrac}[1]{\left(#1\right)}
\newcommand{\ten}[1]{10$^{#1}$}
\newcommand{\tten}[1]{\ifmmode \times10^{#1} \else $\times10^{#1}$\fi}

\newcommand{\zCMB}[1]{\ifmmode z_\mathrm{CMB}^\mathrm{#1} \else $z_\mathrm{CMB}^\mathrm{#1}$\fi}
\newcommand{\zobs}{\ifmmode z_\mathrm{obs} \else $z_\mathrm{obs}$\fi}
\newcommand{\zSun}[1]{\ifmmode z_\mathrm{p,Sun}^\mathrm{{#1}} \else $z_\mathrm{p,Sun}^\mathrm{{#1}}$\fi}
\newcommand\zhelio{\ifmmode z_\mathrm{helio} \else $z_\mathrm{helio}$\fi}
\newcommand\dg{\ifmmode ^{\circ} \else $^{\circ}$\fi}
\newcommand\vSun[1]{\ifmmode v_\mathrm{p,Sun}^\mathrm{{#1}} \else $v_\mathrm{p,Sun}^\mathrm{{#1}}$\fi}
\newcommand{\vrec}{v_{\rm r}}
\newcommand{\vp}{v_{\rm p}}
\newcommand{\vt}{v_{\rm t}}
\newcommand{\zp}{z_{\rm p}}

\shorttitle{Redshift Calculations in NED}
\shortauthors{Carr et al.}

\begin{document}

\title{Improving NASA/IPAC Extragalactic Database Redshift Calculations}

\correspondingauthor{Anthony Carr}
\email{anthony.carr@uq.net.au}

\author[0000-0003-4074-5659]{Anthony Carr}
\affiliation{School of Mathematics and Physics, The University of Queensland, QLD Australia 4072}

\author[0000-0002-4213-8783]{Tamara Davis}
\affiliation{School of Mathematics and Physics, The University of Queensland, QLD Australia 4072}

\begin{abstract}
The NASA/IPAC Extragalactic Database (NED) is an impressive tool for finding near-exhaustive information on millions of astrophysical objects.
Here, we outline a small systematic error that occurs in NED because a low-redshift approximation is used when making the correction from redshifts in the heliocentric frame to the cosmic microwave background (CMB) rest frame.
It means that historically NED systematically misreported the values of CMB-frame redshifts by up to $\sim10^{-3}z$ (about 0.001 at redshift of 1).  
This is a {\em systematic} error, and therefore the impact on applications requiring precise redshifts has the potential to be significant---for example, a systematic redshift error of $\sim10^{-4}$ at {\em low} redshift could resolve the Hubble tension.
We have consulted with the NED team and they are updating the software to remove this systematic error so these corrections are accurate at all redshifts.
Here, we explain the changes and how they impact the redshift values NED currently reports.
\end{abstract}


\section{Introduction} 
\label{sec:intro}
Redshifts are absolutely fundamental to cosmology, and have been so ``easy'' to precisely measure for so long \citep[e.g.,][]{slipher1921,zwicky1933,tonry1979} that the uncertainties in redshifts are often neglected, certain approximations generally remain in use and some forms of potential systematic error remain unaccounted for \citep{wojtak2015,davis2019}.
As other measurements become more precise, systematic errors in redshift are becoming more relevant.
Perhaps the most topical and applicable problem we face in cosmology is the tension in the Hubble constant ($H_0$) between local and global measurements \citep{freedman2019,riess2019,planck2018,wong2020}.
Systematic errors in redshift as small as \ten{-4} could go partway in reconciling (or worsening) the $H_0$ controversy \citep{davis2019} since very small systematic redshift offsets---especially if they occur at low redshift---can have a large impact on the inferred cosmological parameters.

The NASA/IPAC Extragalactic Database (NED)\footnote{\url{https://ned.ipac.caltech.edu/}} keeps track of many redshifts used for cosmology.
In this paper we describe improvements to the computations needed to remove a systematic error in the reporting of redshifts corrected to the cosmic microwave background (CMB) reference frame.
We have also worked with NED staff to update the help pages so those performing heliocentric corrections with NED or other data are aware of the correct method that is accurate at all redshifts.
While the error is small and will have been negligible for most applications \citep[e.g.][]{glanville2020}, it is important to correct since systematic effects could propagate out to inferred cosmological parameters \citep{colless1995, wojtak2015,calcino2017,chaves-montero2018,steinhardt2020}.

We begin by clarifying the theory of redshifts and velocities, and discussing which dipole should be used for heliocentric-to-CMB corrections (Section~\ref{sec:theory}).
We then provide plots of NED redshifts before and after correction to demonstrate the size of the systematic errors (Section~\ref{sec:NEDredshifts}).

\section{Theory}
\label{sec:theory}
NED currently uses an additive approximation to combine different sources of redshift.
In this approximation, \zobs{} is treated as an additive combination of the redshift due to cosmological expansion \zCMB{} and the redshift due to our Sun's (or more correctly the solar system's barycentre) peculiar motion \zSun{} with respect to the CMB. 
When this approximation is used we give \zCMB{} the superscript `$+$', so 
\begin{equation} \label{eq:plus}
    \zobs=\zCMB{+}+\zSun{}.
\end{equation} 
However, the correct way to combine redshifts is to multiplicatively combine factors of $(1+z)$.
This is due to the definition of redshift $(1+z)$ as the ratio of observed to emitted wavelength \citep[see, e.g.,][Section 2.3 for an explanation]{davis2019}.
When this precise equation is used we give \zCMB{} the superscript `$\times$', so
\begin{equation} \label{eq:mult}
    (1+\zobs)=(1+\zCMB{\times})(1+\zSun{}).
\end{equation} 
Expanding this second equation out and subtracting Eq.~(\ref{eq:plus}), 
\begin{equation}
    \zCMB{+}=\zCMB{\times}+\zCMB{\times}\zSun{}.
\end{equation}  
Thus, the difference between using the additive approximation and the correct
multiplicative equation is exactly $\zCMB{\times}\zSun{}$.
A summary of symbols can be found in Table~\ref{tab:nomenclature}.  

The redshift due to our Sun's motion of about $369.82\pm0.11$ \kms{} \citep{planck2018} peaks in the direction of motion at $\zSun{}=0.0012336\pm0.0000003$.
Therefore at low redshift, when $\zCMB{\times}$ and the approximate $\zCMB{+}$ are small, their difference is negligible. By $z\sim 1$, however, the error is on the order of \ten{-3}, which is an order of magnitude larger than most reported statistical errors in spectroscopic redshifts.
The systematic error remains even when using photometric redshifts, which generally have uncertainties $\gtrsim 10^{-3}$---an underlying systematic error by its definition will not be mitigated by averaging over many measurements.

Thankfully, despite this error growing larger with redshift, the impact on $H_0$ measurements from standard candles shrinks; these measurements are more strongly impacted by low-redshift systematic errors \citep{wojtak2015,davis2019}.
Nevertheless, it remains important to remove entirely as it will impact other measurements such as whether dark energy is time varying.

\begin{deluxetable*}{lcp{10cm}}
\tabletypesize{\footnotesize}
\tablenum{1}
\tablecaption{Definitions of Symbols and Their Physical Description.} \label{tab:nomenclature}
\tablewidth{0pt}
\tablehead{\colhead{Symbol} & \colhead{Name} & \colhead{Description}}
\decimalcolnumbers
\startdata
Generic recession velocity; redshift        &   $\vrec{}$; \zCMB{}   & Recession velocity (that appears in the Hubble--Lema\^{i}tre law) and redshift due to the expansion of the universe. The recession velocity is related to the recession redshift through Eq.~(\ref{eq:vrec}). \\
Generic peculiar velocity; redshift         &   $\vp{}$; $\zp{}$   & Peculiar velocity and redshift due to motions that deviate from the Hubble flow. Can refer both to our peculiar motion and the peculiar motion of objects we are observing. \\
Observed redshift                           &        \zobs{}       & The redshift of an object in the heliocentric frame, as given by NED. Contains contributions from the object's and our own peculiar motion relative to the CMB, and the object's recession. (Note this is not strictly observed redshift, as the velocity of the Earth relative to the Sun is generally removed.)\\
CMB-frame redshift derived multiplicatively &    \zCMB{\times}     & The redshift of the object after correcting for our peculiar motion relative to the CMB using Eq.~(\ref{eq:mult}). May still have contributions from the object's peculiar motion. This is the correct CMB-frame redshift.\\
CMB-frame redshift derived additively       &     \zCMB{+}         & The approximate redshift of the object after correcting for our peculiar motion relative to the CMB using Eq.~(\ref{eq:plus}). This is a poor approximation at high $z$.\\
Sun's peculiar redshift/velocity            &   \vSun{}; \zSun{}   & An object's peculiar velocity and peculiar redshift  due to the peculiar velocity of the Sun in the direction of the object, relative to the CMB. Note that when \vSun{} is positive, \zSun{} is negative.\\
Sun's velocity toward dipole                &   \vSun{max}         & The magnitude of the Sun's velocity in the CMB frame, related to \vSun{} through Eq.~(\ref{eq:dot}). \\
Separation from dipole                      &       $\alpha$       & The on-sky angular separation of an object from the direction of the CMB dipole. \vspace{3pt} \\ 
\enddata
\tablecomments{In addition to the base symbols, superscripts of `Planck' and `COBE' will often be added when referring to a parameter whose value changes depending on the dipole value used.}
\end{deluxetable*}

\subsection{Velocities} \label{subsec:velocities}

Velocities are a complex topic in cosmology because they are coordinate dependent and we generally use coordinates that violate some of our na\"{i}ve expectations of velocity behavior.
This is in contrast to redshifts, that are observable, dimensionless ratios and thus unambiguous.

The low-redshift approximation $z=v/c$ is appropriate for both peculiar velocities and (small) recession velocities.
However, the two types of velocities are quite different.
Peculiar velocities, $\vp$, are measured with respect to a local Minkowski (special relativistic) frame, and thus the redshift induced by a velocity obeys the special relativistic Doppler shift formula, 
\begin{equation} \label{eq:zSR}
    1+\zp = \sqrt{\frac{1+\vp/c}{1-\vp/c}}.
\end{equation}
On the other hand, the recession velocity that appears in the Hubble--Lema\^{i}tre law, $\vrec$, cannot be described in any inertial frame.
It can be derived from the Friedmann--Lema\^{i}tre--Robertson--Walker metric as the purely expansion  component of the time derivative of proper distance, $D$ \citep{davis2004}.
The other component, from changes in comoving coordinate caused by gravitation, are the peculiar velocities above.
Indeed for $\vrec=H_0D$ to be true at any $D$, $\vrec$ must be allowed to exceed the speed of light.
(An analogous situation occurs  with superluminal infall velocities of ``space'' inside a black hole, which is why light cannot escape.)
This is a standard result, which is a requirement for homogeneity and isotropy, and does not violate relativity.
Thus recession velocities do not obey the special relativistic Doppler shift formula, but are instead calculated by integrating over the expansion rate of the universe during the photon's propagation \citep{harrison1999}, 
\begin{equation} \label{eq:vrec}
    \vrec=c\int_0^{\zCMB{}}\frac{\mathrm{d}z}{E(z)},
\end{equation}
where the dimensionless Hubble parameter $E(z)=H(z)/H_0=\sqrt{\sum\Omega_i(1+z)^{3(1+w_i)}}$  \citep{peebles1993, ryden2003} with $H(z)$ being the Hubble parameter at the time of emission from a galaxy at redshift $z$, $\Omega_i$ being the normalized density of component $i$ of the universe, and $w_i$ being that component's equation of state.
It is interesting to note that the recession velocity as a function of \textit{redshift} is actually independent of $H_0$. 

The total velocity, $\vt$, of an object moving away from us can be accurately calculated as $\vt = \vrec+\vp$ \citep{davis2004}.
The common additive approximation for redshifts comes from dividing this equation by $c$ and applying the low-$z$ approximation $z=v/c$.
However, given the current precision of other measurements, this will be a poor approximation for any redshift higher than $z\sim0.01$.

\subsection{Converting Heliocentric Redshifts to the CMB Frame} \label{subsec:frame}

To correct \zobs{} to the CMB frame requires finding \zSun{}.
The motion of the solar system adds a peculiar redshift to every object.
That peculiar redshift is highest for objects that are in or directly opposite the direction of motion, resulting in a maximum \zSun{} of order \ten{-3}.
The effective radial peculiar velocity of any object is the projection of the Sun's velocity vector on the position vector of the object,
\begin{equation} \label{eq:dot}
    \vSun{} = \textit{\textbf{v}}_{\text{p,Sun}}\cdot\hat{\textit{\textbf{n}}}_{\text{obj}} = \vSun{max}\cos\alpha,
\end{equation}
where $\vSun{}$ represents the projection of the Sun's peculiar velocity along the line of sight to the object, $\hat{\textit{\textbf{n}}}_{\text{obj}}$ is the object's position vector, and $\alpha$ is the angle separating the dipole direction and the object.\footnote{The formula for the angular separation, $\alpha$, in terms of galactic longitude and latitude, $(l,b)$, is
\begin{equation}
    \cos\alpha= \sin(b)\sin(b_0)+\cos(b)\cos(b_0)\cos(l-l_0), 
\end{equation}
where $(l_0,b_0)$ correspond to the dipole direction.
An equivalent calculation that is more numerically stable at very small and large separations is the Vincenty formula \citep{vincenty1975}, for which we use \textit{Astropy}'s SkyCoord implementation.
Setting $\Delta l=l-l_0$, it reads as
\begin{equation} 
    \tan\alpha = \frac{\sqrt{\left(\cos b_0 \sin \Delta l \right)^2 + \left(\cos b\sin b_0 - \sin b \cos b_0 \cos \Delta l \right)^2}}{\sin b \sin b_0  + \cos b \cos b_0 \cos \Delta l }.
\end{equation}
}

We define both recession and peculiar velocities to be positive when they are moving away from us.
Note that in the case of our own motion, a positive velocity gives a negative redshift (blueshift) because we are moving toward the object being observed.

Since the Sun's velocity is small (order of \ten{2} \kms{}) compared to $c$, the low-$z$ approximation $\zSun{} \approx -\vSun{}/c$ can be used.
However, there is negligible computational disadvantage to using the full special relativistic calculation Eq.~(\ref{eq:zSR}),  
\begin{equation}    
    \zSun{} = \sqrt{\frac{1+(-\vSun{})/c}{1-(-\vSun{})/c}}-1.
\end{equation} 
The minus signs before $\vSun{}$ have been left explicit to ensure that a zero separation ($\alpha=0$ in Eq.~(\ref{eq:dot})) results in the object appearing blueshifted due the our positive velocity toward it.
For the purposes of the plots and equations in this paper, we emphasize that we use the approximation purely for clarity and so our comparisons with NED are cleaner (since it makes negligible difference relative to the size of the other discrepancies).

\begin{figure*}[t!]
    \centering
        \includegraphics[clip, trim={0, 15cm, 0, 15cm}, width=0.75\textwidth]{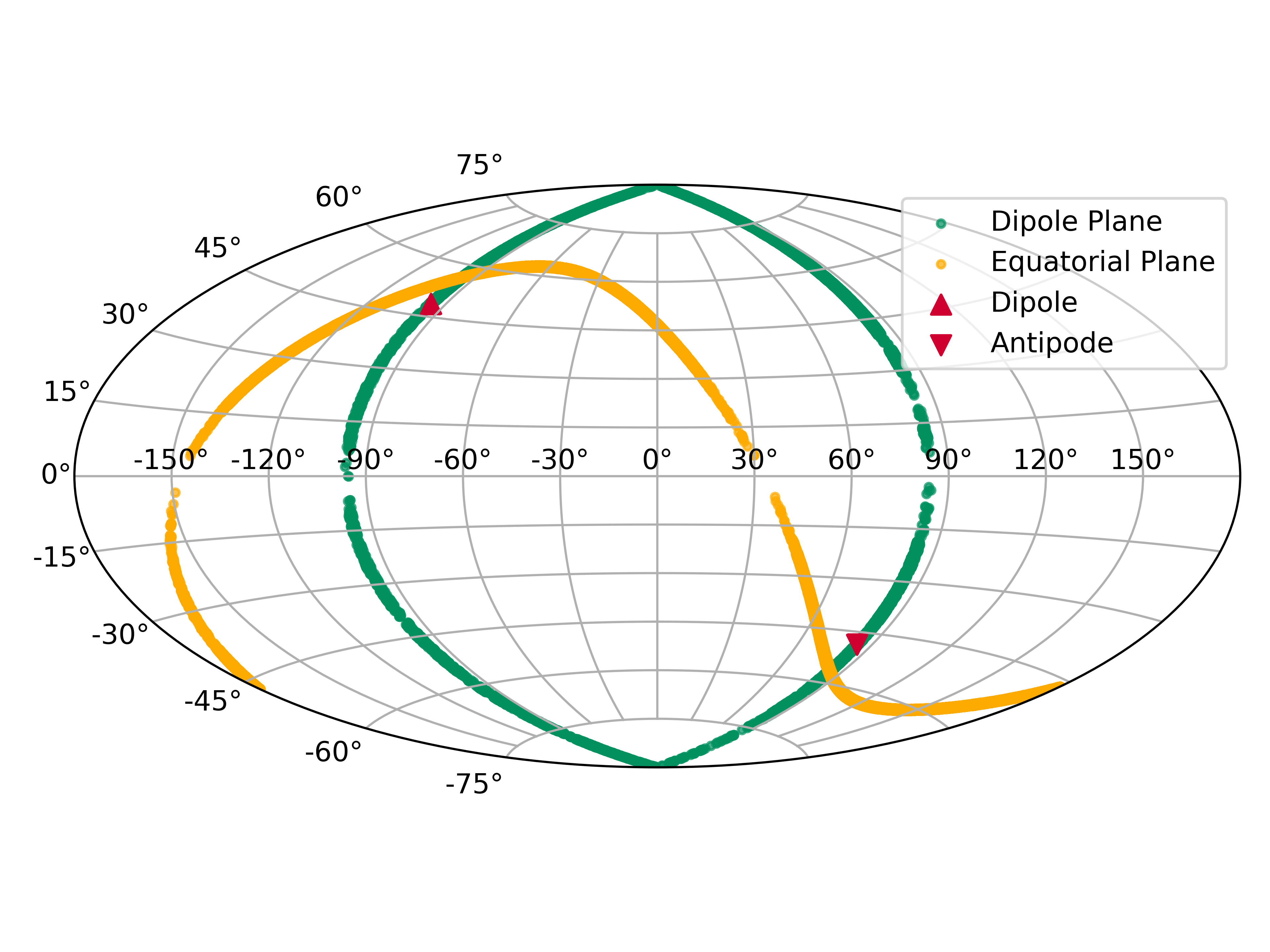}
    \caption{Data sample studied here, showing the equatorial subsample (yellow) and dipole plane (green) in galactic coordinates with the CMB dipole direction and its antipode marked with red triangles.
    }\label{fig:data}
\end{figure*}

\subsection{Multiple CMB Dipole Measurements} \label{subsec:dipole}

The correction requires the direction and magnitude of the CMB dipole, which was measured in 2018 by the Planck Collaboration \citep{planck2018} to be in the direction of $(l, b) = (264.021\dg{}\pm0.011\dg{}, 48.253\dg{}\pm0.005\dg{})$ with a velocity $\vSun{Planck} = 369.82\pm0.11$ \kms{}.
However, NED currently uses the older measurement of the dipole by the Cosmic Background Explorer (COBE) satellite \citep{fixsen1996}, $(l, b) = (264.14\dg{}\pm0.30\dg{}, 48.26\dg{}\pm0.30\dg{})$ with a velocity $\vSun{COBE}= 371\pm1$ \kms{}.\footnote{\url{https://ned.ipac.caltech.edu/Documents/Guides/Calculators}}

We compare the different combinations of dipole and heliocentric correction form below.
Performing the multiplicative correction, Eq.~(\ref{eq:mult}), with the modern dipole measurement is naturally the most accurate.
However, we show explicitly NED uses the additive expansion, Eq.~(\ref{eq:plus}), and the COBE dipole.
In doing so, we demonstrate the size of the resulting systematic errors currently in NED, as well as in the cases of only updating the heliocentric correction form or dipole separately.

\section{Testing NED Redshifts} \label{sec:NEDredshifts}

To test NED redshifts we examine two different data selections.
The first takes objects on a great circle encompassing the dipole direction and its antipode (1.5\dg{} wide), and the second samples the equatorial plane (0.6\dg{} wide), see Fig.~\ref{fig:data}.
As we detail below, adding an equatorial sample is helpful for understanding how errors in the heliocentric correction behave in different parts of the sky.

The combined data set was retrieved using NED in five parts, by splitting the dipole sample into two parts and the equatorial plane into three, for a total of nearly 197,000 objects.
The equatorial sample had to be split into more parts to avoid the queries timing out, as there were nearly an order of magnitude more objects than in the dipole sample.
At the time of retrieval, two XML VOTable files each were required for each data part; one contained the measured parameters, including \zobs{}, and the other contained derived data such as \zCMB{}, to compare to our calculations.
The search criteria in addition to sky position included that the object be a NED classified extragalactic object---excluding {\em Galaxy Groups} and {\em QSO Groups}---with an available redshift.
Despite requesting only extragalactic objects, some objects with negative redshift contaminated the sample.
However, this does not pose a problem since all equations are equally valid for negative redshifts (even if the objects are too local to apply the heliocentric correction).

At the time of writing, the object search process has been altered so that only heliocentric velocities and redshifts can be queried in bulk.

\subsection{Analytic Calculations Compared to Data} \label{subsec:analytical}

\begin{figure*}
    \centering
        \includegraphics[width=0.49\textwidth]{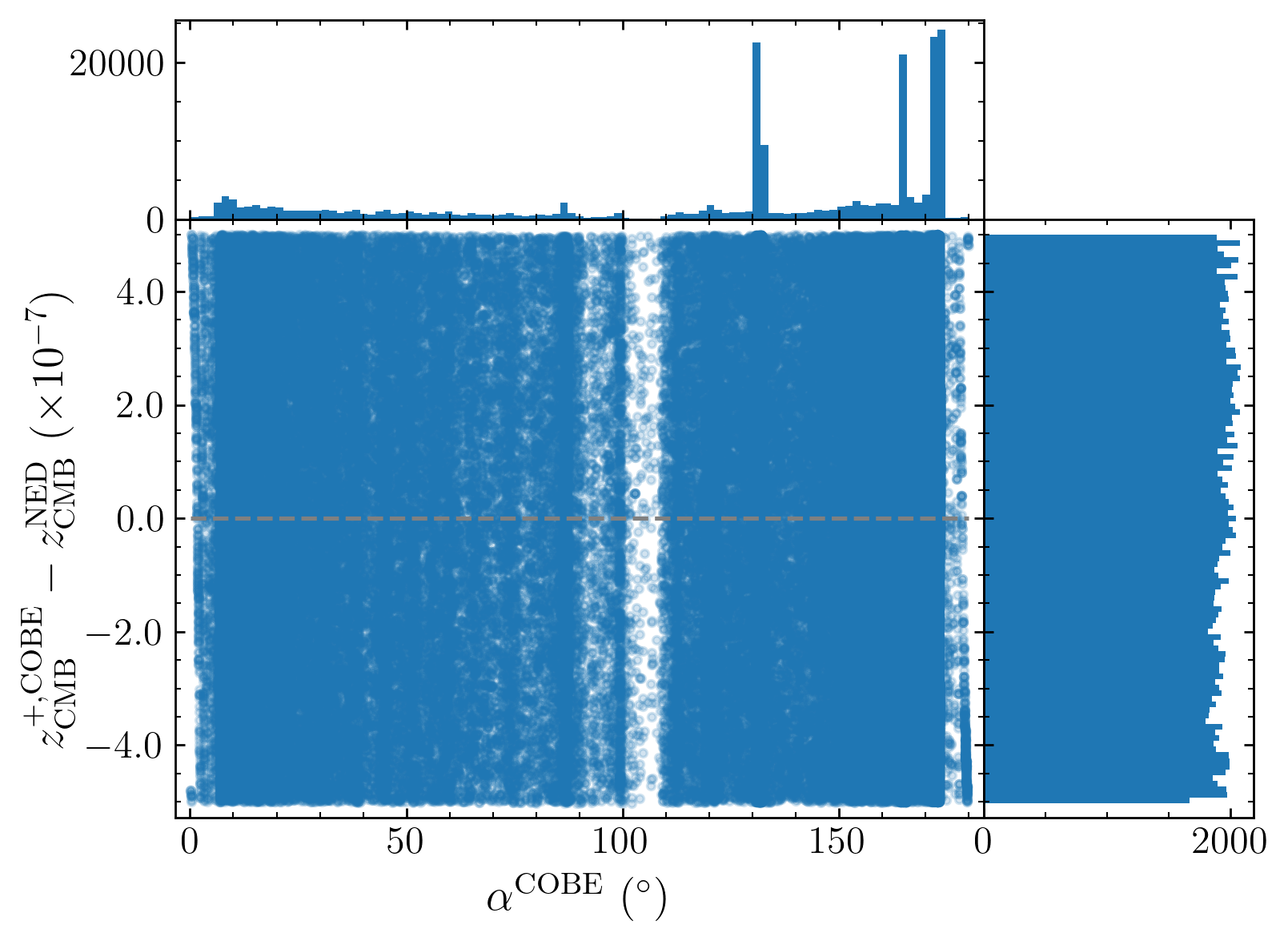}
    \hfill
        \includegraphics[width=0.49\textwidth]{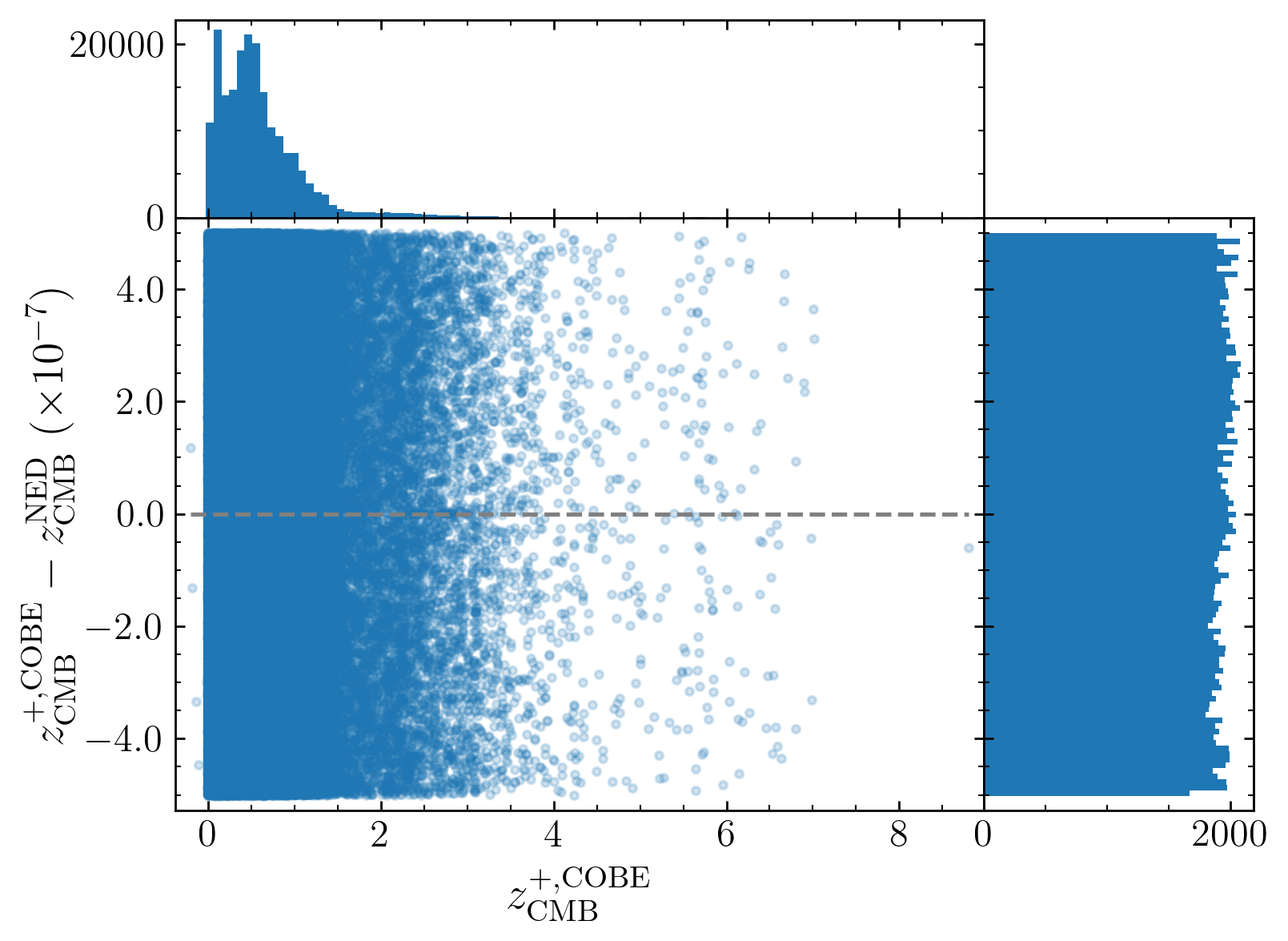}
    \caption{Difference between \zCMB{} calculated using the additive equation with the COBE dipole and the CMB redshift reported in NED. The agreement supports that $\zCMB{NED}=\zCMB{+,COBE}$ since the error arises purely from rounding. \textbf{Left}: the only trend is the clustering at certain separations caused by dense survey regions. \textbf{Right}: along the horizontal axis, the expected distribution of observed galaxies with redshift is evident.}\label{fig:NegligibleDiff}
\end{figure*}

\begin{figure*}
    \centering
        \includegraphics[width=0.49\textwidth]{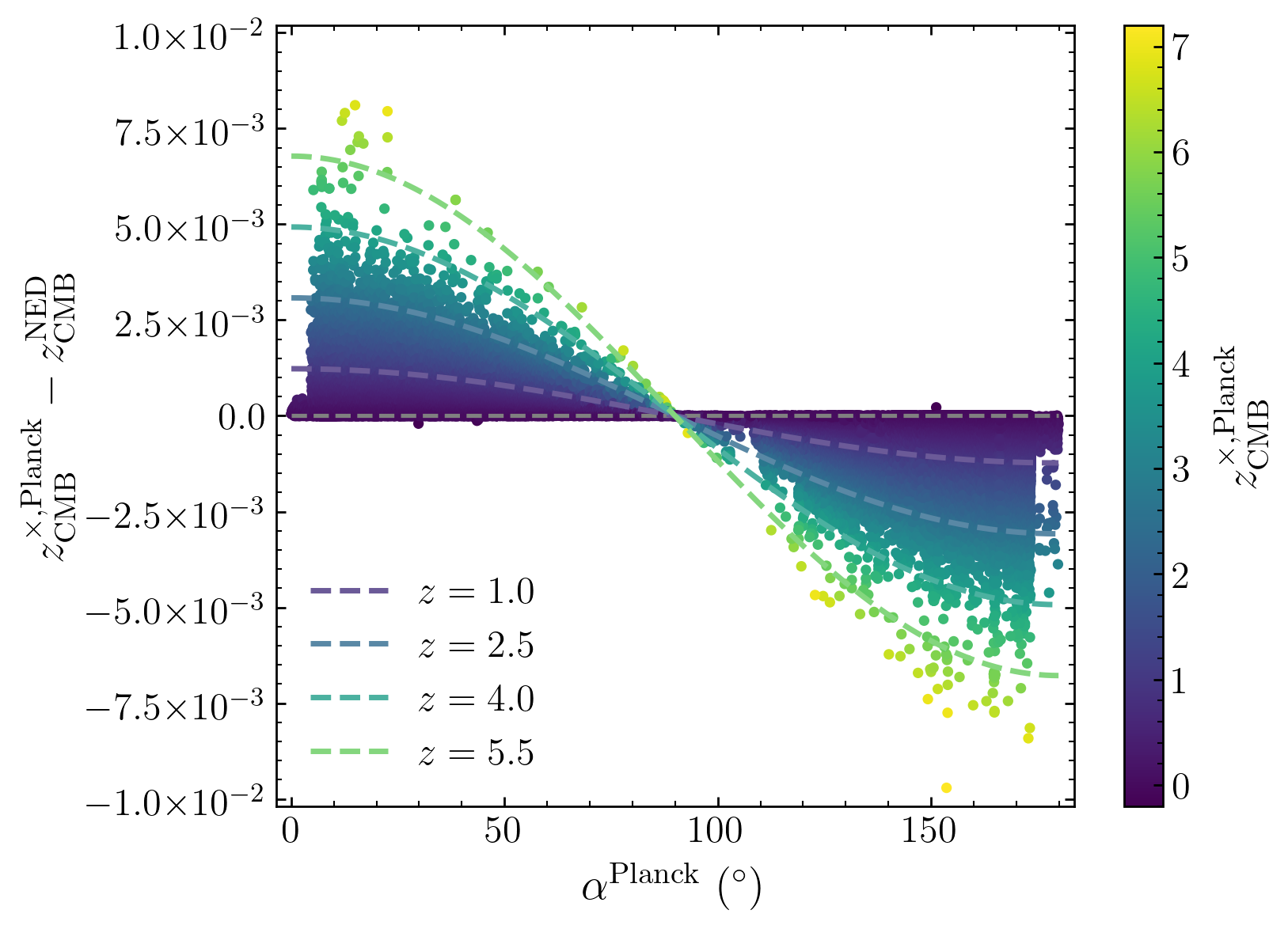}
    \hfill
        \includegraphics[width=0.49\textwidth]{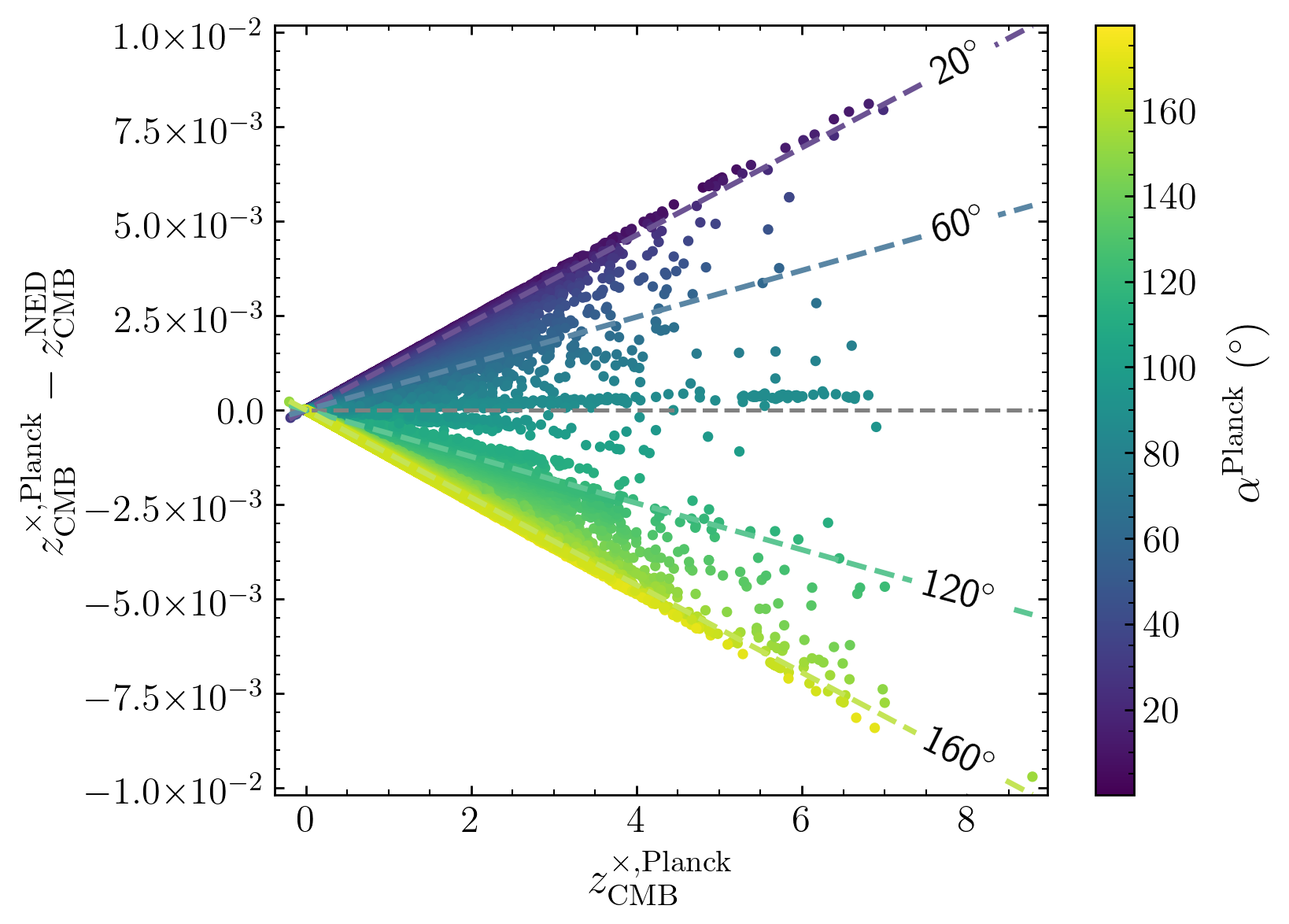}
    \caption{Difference between \zCMB{} calculated using the correct multiplicative equation for the Planck dipole and the CMB redshift reported in NED.
    The lines show the analytic prediction (Eq.~(\ref{eq:multPlanckCOBE})) for the difference between the full calculation and NED values.
    \textbf{Left}: as a function of the objects' separation from the dipole direction, with the color bar representing the CMB-frame redshift.
    \textbf{Right}: as a function of the CMB redshift, with colors representing the angular separation from the dipole direction.
    (The contour colors have been lightened so as to be visible over the data.) }\label{fig:CorrP}
\end{figure*}

\begin{figure*}
    \centering
        \includegraphics[width=0.49\textwidth]{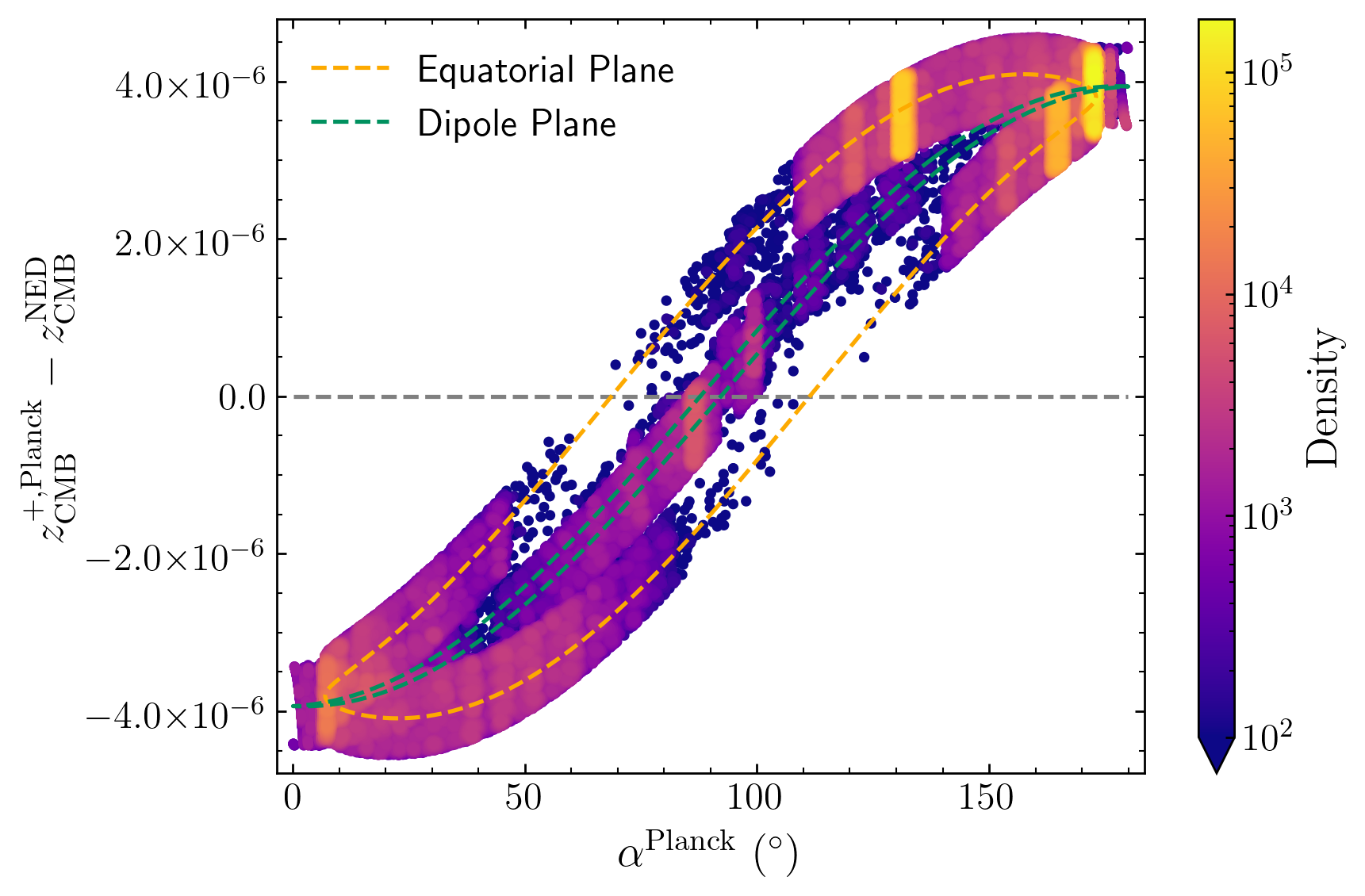}
    \hfill
        \includegraphics[width=0.49\textwidth]{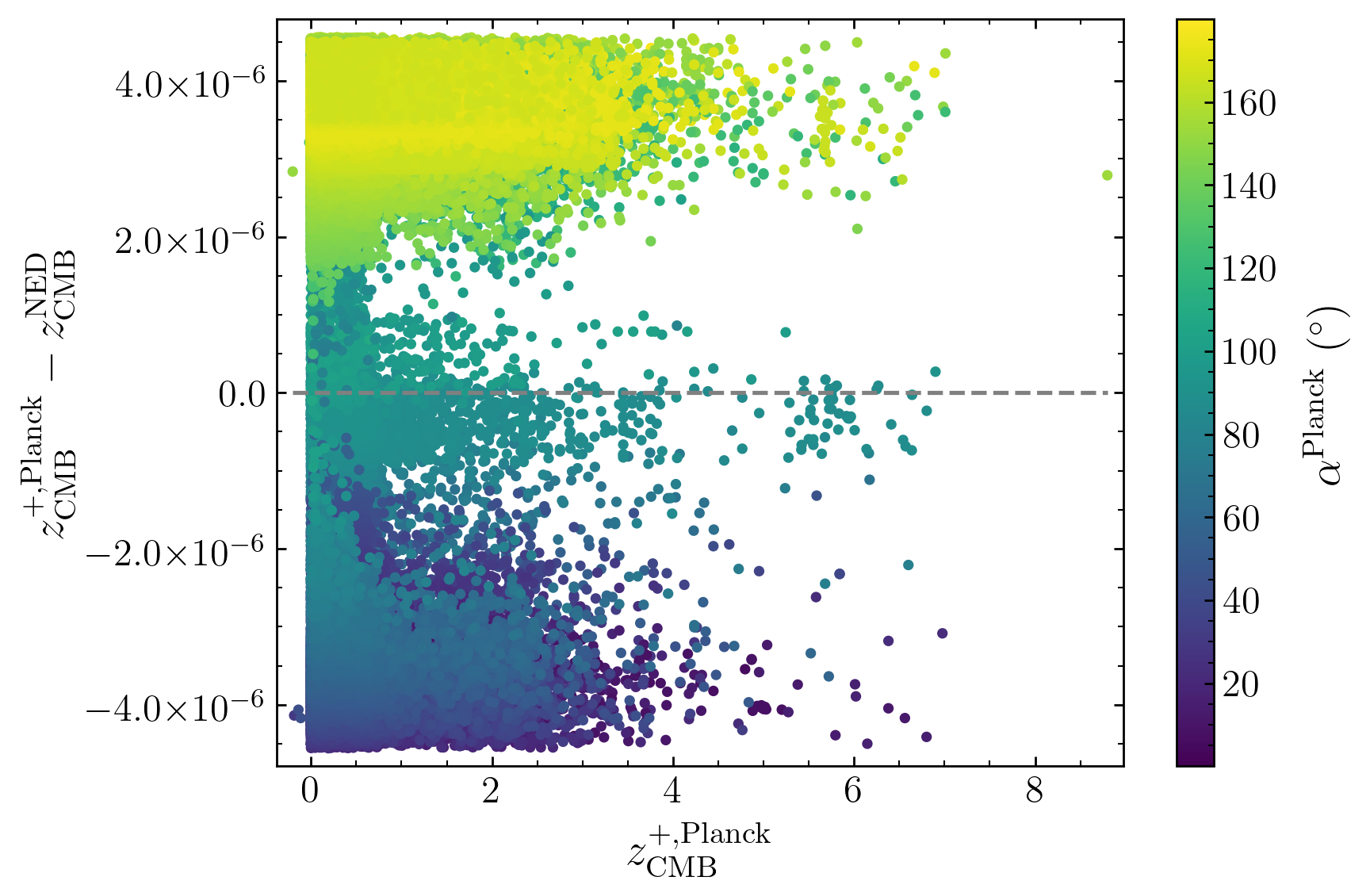}
    \caption{Difference in \zCMB{} using additive+Planck and NED.
    The expected error contours (dashed lines) now depend on exact sky position (the equatorial and dipole samples are now separated) as opposed to just 1D separation.
    \textbf{Left}: as a function of separation from the Planck dipole, with color now representing density of points to contrast the two data samples.
    \textbf{Right}: as a function of calculated CMB redshift, with color representing separation from the Planck dipole.
    The clear trend with separation angle (color bar) has been washed out slightly by the equatorial sample where maximum absolute no longer occurs at extremal (Planck) separations.}\label{fig:IncorrP}
\end{figure*}

\begin{figure*}
    \centering
        \includegraphics[width=0.49\textwidth]{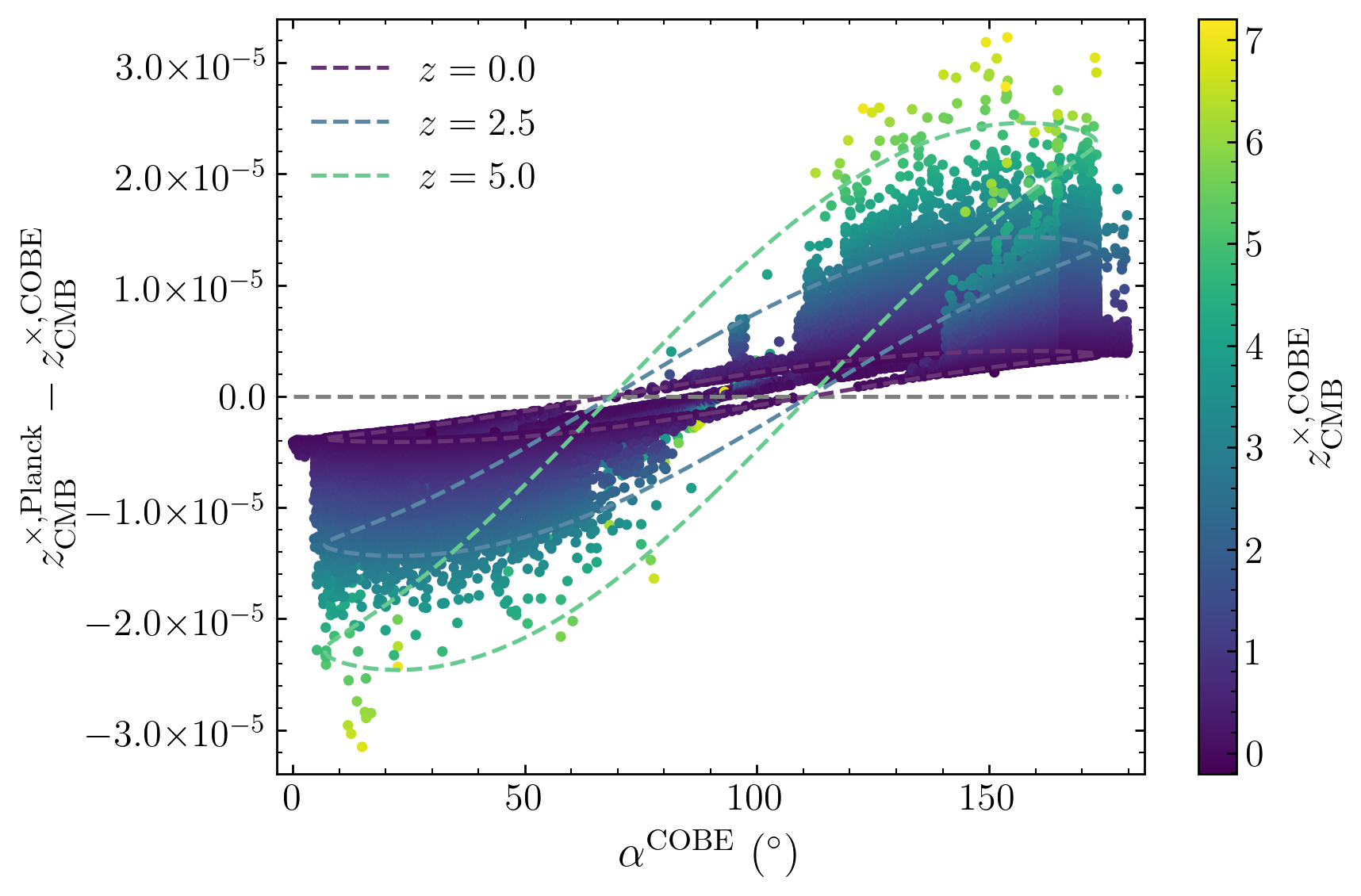}
    \hfill
        \includegraphics[width=0.49\textwidth]{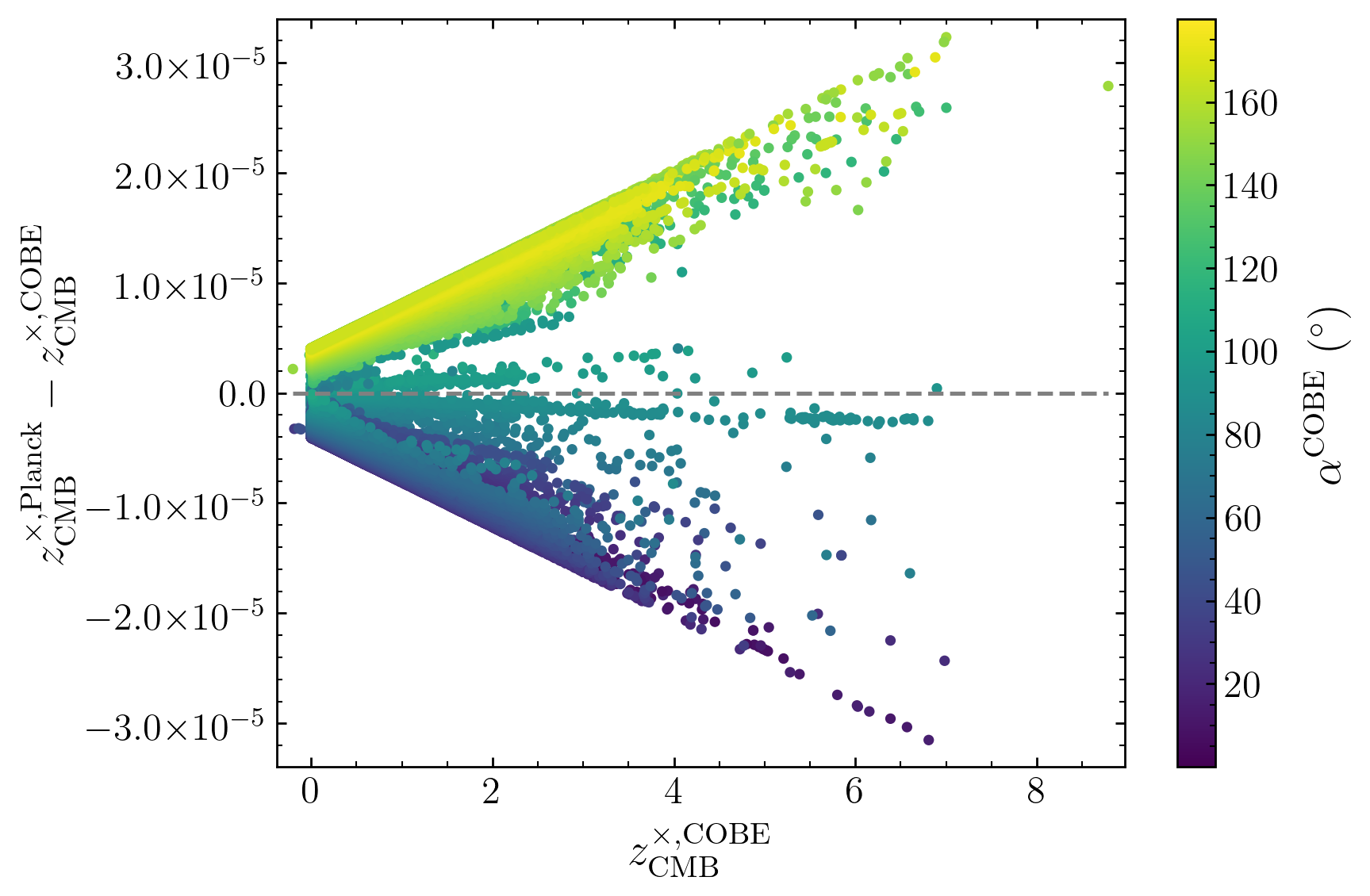}
    \caption{Comparison of full multiplicative heliocentric correction with the two dipoles.
    As discussed in the text, this essentially only differs from Fig.~\ref{fig:IncorrP} by a factor of \zCMB{}.
    \textbf{Left}: as a function of separation.
    There is excellent agreement with the theoretical contours (dashed lines).
    Note that the contours here are only for the equatorial sample.
    \textbf{Right}: as a function of \zCMB{} using the COBE dipole.}\label{fig:CorrF}
\end{figure*}

In Figs.~\ref{fig:NegligibleDiff}--\ref{fig:CorrF}, we compare the CMB redshifts from NED to the CMB redshifts we calculate.
We begin by confirming that NED currently uses the additive redshift calculation and the COBE dipole.
In Fig.~\ref{fig:NegligibleDiff}, we show there is negligible difference in $\zCMB{+,COBE} - \zCMB{NED}$ since the absolute error only reaches 5\tten{-7}.
This error comes purely from comparing the NED redshifts, that are rounded to six decimal places, with our non-rounded calculations.

We then improve on the NED redshifts by using the full multiplicative heliocentric correction and the Planck dipole.
Using the multiplicative redshift calculations results in a redshift difference between our calculations and NED's on the order of \ten{-3}(Fig.~\ref{fig:CorrP}), while changing the dipole results in a redshift difference on the order of \ten{-6}--\ten{-5} (Figs.~\ref{fig:IncorrP} and \ref{fig:CorrF}).
Correcting the current NED CMB redshift calculation is therefore the most important change, while updating the COBE dipole is a valuable but subtle change.

Figure~\ref{fig:CorrP} shows the full difference between the Planck dipole with multiplicative redshift calculation, and the COBE dipole with additive redshift calculation.
That is the difference between Eq.~(\ref{eq:mult}) and Eq.~(\ref{eq:plus}) that we calculate analytically to be
\begin{equation}\label{eq:multPlanckCOBE}
    \zCMB{\times,Planck}-\zCMB{+,COBE} = -\zSun{Planck}\rbrac{1+\zCMB{\times,Planck}}+\zSun{COBE}.
\end{equation} 
Recall that the dipoles differ in both magnitude and direction so there are the two separate expressions for the Sun's peculiar redshift $c\zSun{Planck}\approx -\vSun{max,Planck}\cos\alpha^{\rm Planck}$, and similar for COBE.
We show in Fig.~\ref{fig:CorrP} these analytic calculations (Eq.~(\ref{eq:multPlanckCOBE})) by varying either \zCMB{} or $\alpha$, as dashed lines.
The excellent agreement when replacing \zCMB{NED} with \zCMB{+,COBE} confirms that indeed $\zCMB{NED}=\zCMB{+,COBE}$.

Figure~\ref{fig:IncorrP} highlights only the difference caused by using the Planck dipole over COBE.
For the purposes of this demonstration we calculate the CMB redshift using the additive approximation, but with the Planck dipole.
The difference should therefore be
\begin{equation}\label{eq:IncorrP} 
    \zCMB{+,Planck}-\zCMB{+,COBE} = -\zSun{Planck}+\zSun{COBE}.
\end{equation}
The more complicated shape shown in the left panel of Fig.~\ref{fig:IncorrP} is caused by this difference in two cosines (the different dipole directions) and very clearly shows the difference between the two data samples.
The dipole plane sample has a small variation about the central sinusoid since the Planck dipole is very close to the COBE dipole, but the slight offset of roughly 0.1\dg{} causes a single value of $\alpha^\mathrm{Planck}$ to correspond to two slightly different $\alpha^\mathrm{COBE}$.
The equatorial sample has the larger variation because closest sky position to the dipoles is 6$^{\circ}$ away.

Figure~\ref{fig:CorrF} again compares the two dipoles, but this time using the correct multiplicative redshift calculation for both, 
\begin{eqnarray} 
\zCMB{\times,Planck}-\zCMB{\times,COBE}  =& -\zSun{Planck}\rbrac{1+\zCMB{\times,Planck}} \nonumber \\
& +\zSun{COBE}\rbrac{1+\zCMB{\times,COBE}}. \nonumber \\ &
\end{eqnarray}
Comparing $\zCMB{\times,Planck}$ to \zCMB{NED} would be only imperceptibly different from Fig.~\ref{fig:CorrP}, so instead we opt to show the difference purely caused by the dipoles.
This is essentially Fig.~\ref{fig:IncorrP} and Eq.~(\ref{eq:IncorrP}) with a multiplicative $1+\zCMB{}$ factor.
Therefore, using the full heliocentric correction with the outdated dipole measurement actually also multiplies the error.
Note, however, that the sign of error when comparing dipoles is opposite to the additive versus multiplicative correction.
The increased error here then cancels slightly more with the heliocentric correction form error (albeit still two orders of magnitude smaller).

\section{Conclusion} \label{sec:conclusion}

The additive approximation to calculate the cosmological redshift was reasonable in the past when the correction was first implemented in NED, but this is no longer the case with the high redshifts of modern data.
At a redshift of only 0.1, the error in correcting heliocentric redshifts to the CMB reference frame using the additive approximation
is on the order of \ten{-4}, and this increases linearly with redshift (Fig.~\ref{fig:CorrP}).

A systematic error at that level would noticeably affect standard candle derivations of $H_0$ if it were at low redshift.
The $H_0$ measurements are not likely to be strongly impacted, however, both because this correction is small at low $z$ and because standard candle measurements come from many directions on the sky, so the positive and negative errors tend to cancel out.
However, surveys that cover a small area of the sky are more likely to be biased by this effect.
High $z$ supernova surveys, that tend to monitor small patches of sky, will be more significantly affected (both because of their high redshifts and the narrow $\alpha$ range).
Correcting this is therefore most important for measuring features such as the equation of state of dark energy.  

Finally, we acknowledge that the additive equations and COBE dipole are still used by many in the broader community beyond NED.  We recommend dispensing with these approximations and using the full multiplicative equation to calculate CMB-frame redshifts, while also using the up-to-date Planck dipole. 
These are trivial changes to implement but has the potential to avoid significant errors in resulting analyses.

\acknowledgments
We thank Barry Madore for suggesting we work with the NED team to resolve this small systematic, and Joseph Mazarella and Xiuqin Wu for reviewing this work and overseeing the implementation of the improved redshift corrections by the NED team.
This research has made use of NED, which is funded by the National Aeronautics and Space Administration and operated by the California Institute of Technology.
T.M.D.~is the recipient of an Australian Research Council Australian Laureate Fellowship (grant number FL180100168) funded by the Australian Government.
\software{Astropy v4.0.1 \citep{astropy2018}.}

\bibliography{references}{}

\begin{thebibliography}{}
\expandafter\ifx\csname natexlab\endcsname\relax\def\natexlab#1{#1}\fi
\providecommand{\url}[1]{\href{#1}{#1}}
\providecommand{\dodoi}[1]{doi:~\href{http://doi.org/#1}{\nolinkurl{#1}}}
\providecommand{\doeprint}[1]{\href{http://ascl.net/#1}{\nolinkurl{http://ascl.net/#1}}}
\providecommand{\doarXiv}[1]{\href{https://arxiv.org/abs/#1}{\nolinkurl{https://arxiv.org/abs/#1}}}

\bibitem[{{Calcino} \& {Davis}(2017)}]{calcino2017}
{Calcino}, J., \& {Davis}, T. 2017, \jcap, 2017, 038,
  \dodoi{10.1088/1475-7516/2017/01/038}

\bibitem[{{Chaves-Montero} {et~al.}(2018){Chaves-Montero}, {Angulo}, \&
  {Hern{\'a}ndez-Monteagudo}}]{chaves-montero2018}
{Chaves-Montero}, J., {Angulo}, R.~E., \& {Hern{\'a}ndez-Monteagudo}, C. 2018,
  \mnras, 477, 3892, \dodoi{10.1093/mnras/sty924}

\bibitem[{{Colless}(1995)}]{colless1995}
{Colless}, M. 1995, \aj, 109, 1937, \dodoi{10.1086/117419}

\bibitem[{{Davis} {et~al.}(2019){Davis}, {Hinton}, {Howlett}, \&
  {Calcino}}]{davis2019}
{Davis}, T.~M., {Hinton}, S.~R., {Howlett}, C., \& {Calcino}, J. 2019, \mnras,
  490, 2948, \dodoi{10.1093/mnras/stz2652}

\bibitem[{{Davis} \& {Lineweaver}(2004)}]{davis2004}
{Davis}, T.~M., \& {Lineweaver}, C.~H. 2004, \pasa, 21, 97,
  \dodoi{10.1071/AS03040}

\bibitem[{{Fixsen} {et~al.}(1996){Fixsen}, {Cheng}, {Gales}, {Mather},
  {Shafer}, \& {Wright}}]{fixsen1996}
{Fixsen}, D.~J., {Cheng}, E.~S., {Gales}, J.~M., {et~al.} 1996, \apj, 473, 576,
  \dodoi{10.1086/178173}

\bibitem[{{Freedman} {et~al.}(2019){Freedman}, {Madore}, {Hatt}, {Hoyt},
  {Jang}, {Beaton}, {Burns}, {Lee}, {Monson}, {Neeley}, {Phillips}, {Rich}, \&
  {Seibert}}]{freedman2019}
{Freedman}, W.~L., {Madore}, B.~F., {Hatt}, D., {et~al.} 2019, \apj, 882, 34,
  \dodoi{10.3847/1538-4357/ab2f73}

\bibitem[{{Glanville} {et~al.}(2021){Glanville}, {Howlett}, \&
  {Davis}}]{glanville2020}
{Glanville}, A., {Howlett}, C., \& {Davis}, T.~M. 2021, \mnras, 503, 3510,
  \dodoi{10.1093/mnras/stab657}

\bibitem[{{Harrison}(1993)}]{harrison1999}
{Harrison}, E. 1993, \apj, 403, 28, \dodoi{10.1086/172179}

\bibitem[{Peebles(1993)}]{peebles1993}
Peebles, P. J.~E. 1993, Principles of Physical Cosmology, Princeton series in
  physics (Princeton University Press)

\bibitem[{{Planck Collaboration} {et~al.}(2020){Planck Collaboration},
  {Aghanim}, {Akrami}, {Arroja}, {Ashdown}, {Aumont}, {Baccigalupi},
  {Ballardini}, {Banday}, {Barreiro}, {Bartolo}, {Basak}, {Battye}, {Benabed},
  {Bernard}, {Bersanelli}, {Bielewicz}, {Bock}, {Bond}, {Borrill}, {Bouchet},
  {Boulanger}, {Bucher}, {Burigana}, {Butler}, {Calabrese}, {Cardoso},
  {Carron}, {Casaponsa}, {Challinor}, {Chiang}, {Colombo}, {Combet},
  {Contreras}, {Crill}, {Cuttaia}, {de Bernardis}, {de Zotti}, {Delabrouille},
  {Delouis}, {D{\'e}sert}, {Di Valentino}, {Dickinson}, {Diego}, {Donzelli},
  {Dor{\'e}}, {Douspis}, {Ducout}, {Dupac}, {Efstathiou}, {Elsner},
  {En{\ss}lin}, {Eriksen}, {Falgarone}, {Fantaye}, {Fergusson},
  {Fernandez-Cobos}, {Finelli}, {Forastieri}, {Frailis}, {Franceschi},
  {Frolov}, {Galeotta}, {Galli}, {Ganga}, {G{\'e}nova-Santos}, {Gerbino},
  {Ghosh}, {Gonz{\'a}lez-Nuevo}, {G{\'o}rski}, {Gratton}, {Gruppuso},
  {Gudmundsson}, {Hamann}, {Handley}, {Hansen}, {Helou}, {Herranz},
  {Hildebrandt}, {Hivon}, {Huang}, {Jaffe}, {Jones}, {Karakci}, {Keih{\"a}nen},
  {Keskitalo}, {Kiiveri}, {Kim}, {Kisner}, {Knox}, {Krachmalnicoff}, {Kunz},
  {Kurki-Suonio}, {Lagache}, {Lamarre}, {Langer}, {Lasenby}, {Lattanzi},
  {Lawrence}, {Le Jeune}, {Leahy}, {Lesgourgues}, {Levrier}, {Lewis},
  {Liguori}, {Lilje}, {Lilley}, {Lindholm}, {L{\'o}pez-Caniego}, {Lubin}, {Ma},
  {Mac{\'\i}as-P{\'e}rez}, {Maggio}, {Maino}, {Mandolesi}, {Mangilli},
  {Marcos-Caballero}, {Maris}, {Martin}, {Martinelli},
  {Mart{\'\i}nez-Gonz{\'a}lez}, {Matarrese}, {Mauri}, {McEwen}, {Meerburg},
  {Meinhold}, {Melchiorri}, {Mennella}, {Migliaccio}, {Millea}, {Mitra},
  {Miville-Desch{\^e}nes}, {Molinari}, {Moneti}, {Montier}, {Morgante}, {Moss},
  {Mottet}, {M{\"u}nchmeyer}, {Natoli}, {N{\o}rgaard-Nielsen}, {Oxborrow},
  {Pagano}, {Paoletti}, {Partridge}, {Patanchon}, {Pearson}, {Peel}, {Peiris},
  {Perrotta}, {Pettorino}, {Piacentini}, {Polastri}, {Polenta}, {Puget},
  {Rachen}, {Reinecke}, {Remazeilles}, {Renault}, {Renzi}, {Rocha}, {Rosset},
  {Roudier}, {Rubi{\~n}o-Mart{\'\i}n}, {Ruiz-Granados}, {Salvati}, {Sandri},
  {Savelainen}, {Scott}, {Shellard}, {Shiraishi}, {Sirignano}, {Sirri},
  {Spencer}, {Sunyaev}, {Suur-Uski}, {Tauber}, {Tavagnacco}, {Tenti},
  {Terenzi}, {Toffolatti}, {Tomasi}, {Trombetti}, {Valiviita}, {Van Tent},
  {Vibert}, {Vielva}, {Villa}, {Vittorio}, {Wandelt}, {Wehus}, {White},
  {White}, {Zacchei}, \& {Zonca}}]{planck2018}
{Planck Collaboration}, {Aghanim}, N., {Akrami}, Y., {et~al.} 2020, \aap, 641,
  A1, \dodoi{10.1051/0004-6361/201833880}

\bibitem[{{Riess} {et~al.}(2019){Riess}, {Casertano}, {Yuan}, {Macri}, \&
  {Scolnic}}]{riess2019}
{Riess}, A.~G., {Casertano}, S., {Yuan}, W., {Macri}, L.~M., \& {Scolnic}, D.
  2019, \apj, 876, 85, \dodoi{10.3847/1538-4357/ab1422}

\bibitem[{Ryden(2003)}]{ryden2003}
Ryden, B.~S. 2003, Introduction to Cosmology (Addison-Wesley)

\bibitem[{{Slipher}(1921)}]{slipher1921}
{Slipher}, V.~M. 1921, PA, 29, 128

\bibitem[{{Steinhardt} {et~al.}(2020){Steinhardt}, {Sneppen}, \&
  {Sen}}]{steinhardt2020}
{Steinhardt}, C.~L., {Sneppen}, A., \& {Sen}, B. 2020, \apj, 902, 14,
  \dodoi{10.3847/1538-4357/abb140}

\bibitem[{{The Astropy Collaboration} {et~al.}(2018){The Astropy
  Collaboration}, {Price-Whelan}, {Sip{\H{o}}cz}, {G{\\"u}nther}, {Lim},
  {Crawford}, {Conseil}, {Shupe}, {Craig}, {Dencheva}, {Ginsburg},
  {VanderPlas}, {Bradley}, {P{\\\'e}rez-Su{\\\'a}rez}, {de Val-Borro}, {Paper
  Contributors}, {Aldcroft}, {Cruz}, {Robitaille}, {Tollerud}, {Coordination
  Committee}, {Ardelean}, {Babej}, {Bach}, {Bachetti}, {Bakanov}, {Bamford},
  {Barentsen}, {Barmby}, {Baumbach}, {Berry}, {Biscani}, {Boquien}, {Bostroem},
  {Bouma}, {Brammer}, {Bray}, {Breytenbach}, {Buddelmeijer}, {Burke},
  {Calderone}, {Cano Rodr{\\\'{\\i}}guez}, {Cara}, {Cardoso}, {Cheedella},
  {Copin}, {Corrales}, {Crichton}, {D{\\rsquo}Avella}, {Deil}, {Depagne},
  {Dietrich}, {Donath}, {Droettboom}, {Earl}, {Erben}, {Fabbro}, {Ferreira},
  {Finethy}, {Fox}, {Garrison}, {Gibbons}, {Goldstein}, {Gommers}, {Greco},
  {Greenfield}, {Groener}, {Grollier}, {Hagen}, {Hirst}, {Homeier}, {Horton},
  {Hosseinzadeh}, {Hu}, {Hunkeler}, {Ivezi{\\\'c}}, {Jain}, {Jenness},
  {Kanarek}, {Kendrew}, {Kern}, {Kerzendorf}, {Khvalko}, {King}, {Kirkby},
  {Kulkarni}, {Kumar}, {Lee}, {Lenz}, {Littlefair}, {Ma}, {Macleod},
  {Mastropietro}, {McCully}, {Montagnac}, {Morris}, {Mueller}, {Mumford},
  {Muna}, {Murphy}, {Nelson}, {Nguyen}, {Ninan}, {N{\\"o}the}, {Ogaz}, {Oh},
  {Parejko}, {Parley}, {Pascual}, {Patil}, {Patil}, {Plunkett}, {Prochaska},
  {Rastogi}, {Reddy Janga}, {Sabater}, {Sakurikar}, {Seifert}, {Sherbert},
  {Sherwood-Taylor}, {Shih}, {Sick}, {Silbiger}, {Singanamalla}, {Singer},
  {Sladen}, {Sooley}, {Sornarajah}, {Streicher}, {Teuben}, {Thomas},
  {Tremblay}, {Turner}, {Terr{\\\'o}n}, {van Kerkwijk}, {de la Vega},
  {Watkins}, {Weaver}, {Whitmore}, {Woillez}, {Zabalza}, \&
  {Contributors}}]{astropy2018}
{The Astropy Collaboration}, {Price-Whelan}, A.~M., {Sip{\H{o}}cz}, B.~M.,
  {et~al.} 2018, \aj, 156, 123, \dodoi{10.3847/1538-3881/aabc4f}

\bibitem[{{Tonry} \& {Davis}(1979)}]{tonry1979}
{Tonry}, J., \& {Davis}, M. 1979, \aj, 84, 1511, \dodoi{10.1086/112569}

\bibitem[{Vincenty(1975)}]{vincenty1975}
Vincenty, T. 1975, Survey Review, 23, 88, \dodoi{10.1179/sre.1975.23.176.88}

\bibitem[{{Wojtak} {et~al.}(2015){Wojtak}, {Davis}, \& {Wiis}}]{wojtak2015}
{Wojtak}, R., {Davis}, T.~M., \& {Wiis}, J. 2015, \jcap, 2015, 025,
  \dodoi{10.1088/1475-7516/2015/07/025}

\bibitem[{{Wong} {et~al.}(2020){Wong}, {Suyu}, {Chen}, {Rusu}, {Millon},
  {Sluse}, {Bonvin}, {Fassnacht}, {Taubenberger}, {Auger}, {Birrer}, {Chan},
  {Courbin}, {Hilbert}, {Tihhonova}, {Treu}, {Agnello}, {Ding}, {Jee},
  {Komatsu}, {Shajib}, {Sonnenfeld}, {Blandford}, {Koopmans}, {Marshall}, \&
  {Meylan}}]{wong2020}
{Wong}, K.~C., {Suyu}, S.~H., {Chen}, G. C.~F., {et~al.} 2020, \mnras, 498,
  1420, \dodoi{10.1093/mnras/stz3094}

\bibitem[{{Zwicky}(1933)}]{zwicky1933}
{Zwicky}, F. 1933, AcHPh, 6, 110

\end{thebibliography}

\end{document}